\documentclass{ws-procs961x669}            

\usepackage{graphicx}
\usepackage{latexsym}

\newcommand{\bea}{\begin{eqnarray}}	
\newcommand{\eea}{\end{eqnarray}}
\newcommand{\be}{\begin{equation}}	
\newcommand{\ee}{\end{equation}}
\newcommand{\beq}{\begin{equation}}	
\newcommand{\eeq}{\end{equation}}

\newcommand{\C}{{\mathbb C}}
\newcommand{\Z}{ {\mathbb Z} } 
 
\newcommand{\N}{ {\mathbb N} } 
\newcommand{\cG}{\mathcal{G}}
 
 \newcommand{\cV}{\mathcal{V}}
 
 \newcommand{\cL}{\mathcal{L}}

\newcommand{\cF}{ { \mathcal{ F } } }

\newcommand{\R}{\mathbb{R}} 
\newcommand{\Tr}{{\rm Tr}}

\newcommand{\bdel}{{\boldsymbol{\delta}}}

\newcommand{\inter}{{\rm int\,}}

\begin{document}
\title{Renormalizable Tensor Field Theories\footnote{Based on the talk
``Tensor Models and Renormalization'' given 
at the International Congress on Mathematical Physics, ICMP2015, 
27th July - 1st August, 2015, Santiago de Chile, Chile.}  } 

\author{J. Ben Geloun}

\address{Max-Planck Institute for Gravitational Physics,
Albert Einstein Institute \\
Am M\"uhlenberg 1, Potsdam, D-14476, Germany\\
International Chair in Mathematical Physics and Applications \\
ICMPA-UNESCO Chair, 072Bp50 Cotonou, Rep of Benin \\
E-mail: jbengeloun@aei.mpg.de }

\begin{abstract}
Extending tensor models at the field theoretical level, 
tensor field theories are nonlocal quantum field theories
with Feynman graphs identified with simplicial complexes. 
They become relevant for addressing quantum topology and geometry 
in any dimension and therefore form an interesting class of models for studying quantum gravity. 
We review the class of perturbatively renormalizable tensor field theories and some of their features. 
\end{abstract}


\bodymatter

\section{Introduction}\label{intro}

Introduced soon after the success of matrix models\cite{Di Francesco:1993nw} and aiming at generalizing their success, tensor models\cite{tensor} which address the weighted sum of simplicial  manifolds in any dimension, turned out to be dramatically more involved than their lower dimensional cousins. 
We must recall that, among approaches for quantum gravity in low dimensions, matrix models\cite{Di Francesco:1993nw} remain a prominent framework. Their achievement largely relies on the fact that geometry 
in 2D is of course well understood and on a fundamental tool which has  given an handle 
on the partition function of matrix models: the 't Hooft large 
$N$ expansion. Back in the 90's, much less is known about path integral of tensor models. Their phase transition and resulting geometries which turned out to be singular were investigated only through numerics.  
This approach needed a drastic change and rethinking.
At the same period, Boulatov\cite{boulatov} finds a link
between a field theory formulation of tensor models and 
the Ponzano-Regge model for 3D gravity in the form of a lattice gauge field theory. The model by Boulatov
introduces the concept of group holonomies in cellular complexes associated with Feynman graphs.
It did not take long to see emerging a new framework called Group Field Theory (GFT).\cite{oriti} 

In 2010,  
Gurau discovered\cite{largeN} a large $N$ expansion of a particular class
of tensor models called colored.\cite{coloured} 
Then, a series of results followed among which the analytical proof that
colored tensor models underdo a phase transition.\cite{criticalTensor} After transition the type of geometries in 
the continuum limit was shown  very singular: the so-called branched
polymer geometries. This means that colored tensor
models must be again enriched by other data to fulfill their goal
to describe a large and smooth geometry in the continuum,
 a spacetime with all properties pertaining to it. 

Simply because are quantum field theories, the question of the
renormalization analysis in GFTs has been addressed in
the meantime without a complete answer.\cite{gfttrial} 
In any renormalization program, there are difficulties of controlling the type divergences of amplitudes.  This is the central question of
finding a locality principle in field theory: to which type of interactions 
does correspond which type of propagator  which
will allow to perform the subtraction of any primitively divergent graph?
For GFT models, this question was complex
even for the simplest actions. 
 
With the advent of colored tensor models, new types of  
effective interactions (integrating all colors except one\cite{Gurau:2012ix,Bonzom:2012hw}) could have been
studied under the renormalization group lens.
 A first tensor model in 4D, called Tensor Field Theory, was proved renormalizable at all orders.\cite{BenGeloun:2011rc} 
In the following years, a wealth of renormalizable TFTs has
been revealed. Opening a window on nonlocal  QFTs, the renormalization analysis of TFTs has been performed in several dimensions (3 up to 6), using different background spaces (over Abelian, $U(1)$ and $\R$, and non Abelian $SU(2)$, direct spaces), and on models implementing the gauge constraint of GFT.\cite{GFTrenorm, rankd, BenGeloun:2012pu, Carrozza:2013mna, josephBeta, Sylvain, Benedetti:2014qsa, 
Rivasseau:2015ova} See the review.\cite{Rivasseau:2011hm}

The study of renormalization  covers more than a 
mathematical treatment for curing divergences of any QFT. It must explain the physics behind the model.
In particular, the renormalization group (RG) flow analysis of TFTs should deliver at least hints for obtaining a classical spacetime at low energy in some parameter regime. To that extent, perturbative studies of the RG flow 
of TFTs have been undertaken with interesting results: many models
turns out to be asymptotically free (AF).\cite{BenGeloun:2012pu,Carrozza:2013mna, josephBeta, Sylvain,Rivasseau:2015ova}
Asymptotic freedom has been a striking feature for the theory of 
Quantum Chromodynamics (QCD). Roughly speaking, at very high energy, an AF model flows 
toward a free and well-defined theory, whereas going in lower energy, 
the coupling constant of the model grows. This suggests
a radical change of the theory involving a change of degrees of freedom.  In QCD, the coupling of quarks increases which induces
a binding, or confinement,  of these particles 
which produces Hadrons, the most stable of which form
the atom nucleus.  
 For Tensorial Field Theory, asymptotic freedom becomes interesting
mechanism indeed because we do not want to stay in a phase where the geometrical spacetime is apparently discrete and spanned by  building blocks. The hope here is that, to draw a parallel 
with QCD,  asymptotic freedom will induce 
a new phase for TFT models, towards new degrees of freedom
 able to generate a space with properties close to those
of our spacetime. 

The next section reviews the main ingredients to built a 
renormalizable TFT and section \ref{ccl} gives a summary 
of our work and a future direction for investigations.

\section{Tensor field theories}
\label{sect:TFTs} 

\noindent{\bf TFT models.}
Let $\phi_{\bf P}$ be  a rank $d$ complex tensor, where ${\bf P}=(p_1,p_2,\dots,p_d)$ a collection of indices. We denote $\bar\phi_{\bf P}$ its complex conjugate. In this work and for simplicity, we fix $p_k \in \Z$. A motivation on this choice is simple: introducing a complex function $\phi: U(1)^{d} \to \C$,
$\phi_{\bf P}$ define nothing but the Fourier components of $\phi$. 
In this way, a TFT, as described below, defines nothing but
a particular field theory written in the momentum space 
of the torus. The physics here is given through the following 
{\it duality}: $\phi_{\bf P}$ is viewed as a $(d-1)$-simplex. 

A TFT model is defined via an action $S$ built by convoluting  copies
of $\phi_{\bf P}$ and $\bar\phi_{\bf P}$ using kernels: 
\bea
\label{eq:actiond}
&&
S[\bar\phi,\phi]=\Tr_2 (\bar\phi \cdot K \cdot \phi) 
+ \mu\, \Tr_2 (\phi^2) + S^{\inter}[\bar\phi,\phi]\,, 
\crcr
&&
\Tr_2 (\bar\phi \cdot K \cdot \phi) =
\sum_{{\bf P}, \, {\bf P}'} \bar\phi_{{\bf P}} \, K({\bf P};{\bf P}')\, \phi_{{\bf P}' } \,, 
\qquad 
\Tr_{2}(\phi^2) = \sum_{{\bf P}} \bar\phi_{{\bf P}}\phi_{{\bf P}}\,, 
 \crcr
&& S^{\inter}[\bar\phi,\phi]=  \sum_{n_b} 
\lambda_{n_b}  \Tr_{n_b}(\bar\phi^{n_b}\cdot \cV_{n_b}\cdot \phi^{n_b})\,,
\eea
where $\Tr_{n_b}$ can be thought as generalized traces. Each of these  expresses a type of convolution of the indices of the $n_b$ couples of tensors according
to a graphical pattern $b$ (precisions on this will be given in a moment). 
The kernel  $K$ and $\cV_{n_b}$ are to be specified, $\mu$ is a mass and $\lambda_{n_b}$ an interaction coupling constant. Setting $\cV_{n_b}$ to
be of weight 1 kernel, $\Tr_{n_b}$ generate unitary invariants.\cite{Gurau:2012ix}
Recalling that a rank $d$ tensor is dual to a $(d-1)$-simplex,
the contraction of tensors forming an interaction represents a 
$d$-simplex obtained by gluing the $(d-1)$-simplexes
along their $(d-2)$ boundary simplexes. 

Let us now specify the kinetic term  in a rank $d$ action:
\beq
\label{eq:3dkin}
K(\{p_i\};\{p_i'\}) = \bdel_{p_i,p_i'} ( \sum_{i=1}^d p^{2a}_i)  \,,
\;\,
\bdel_{p_i,p_i'} := \prod_{i=1}^d \delta_{p_i,p_i'} \,,
\;\,
\Tr_{2}(\phi^2)= \sum_{p_i \in \Z} |\phi_{12\dots d} |^2 \,,
\eeq
where $a\in (0,1]$, and where we use the notation $\phi_{12\dots d}:=\phi_{p_1,p_2,\dots, p_d}$. The kernel $K$  is the sum of  $2a$-power eigenvalues of $d$ Laplacian operators acting over the $d$ copies of $U(1)$.
Thus, the dynamics of any model is quite standard. 
We focus now on interactions. To make clear 
the  type of interactions we are considering, let us
restrict to the rank $d=3$, with a tensor $\phi_{123} := \phi_{p_1,p_2,p_3}$ (the general case can be inferred with no issue). 
To be even more specific, let us construct a $\phi^4$-like tensor
field theory by convoluting four tensors. A first remarkable
thing is that there is more than one way of convoluting
 four tensors. One of the possibilities is given by 
\beq
\Tr_{4;1}(\phi^4) = \sum_{p_i, p_i' } 
\phi_{123} \,\bar\phi_{1'23} \,\phi_{1'2'3'} \,\bar\phi_{12'3'}\,.
\label{choice1}
\eeq
 Hence, a $\phi^4$-theory may include several types of interactions
and, naturally, $\phi^{2n>4}$ theories
become  far richer. The particular pattern of
convolution of \eqref{choice1} will be explained graphically in the next paragraph discussing quantum aspects of the theory. 

\begin{figure}[h]
 \centering
     \begin{minipage}{.9\textwidth}
\includegraphics[angle=0, width=5cm, height=1.5cm]{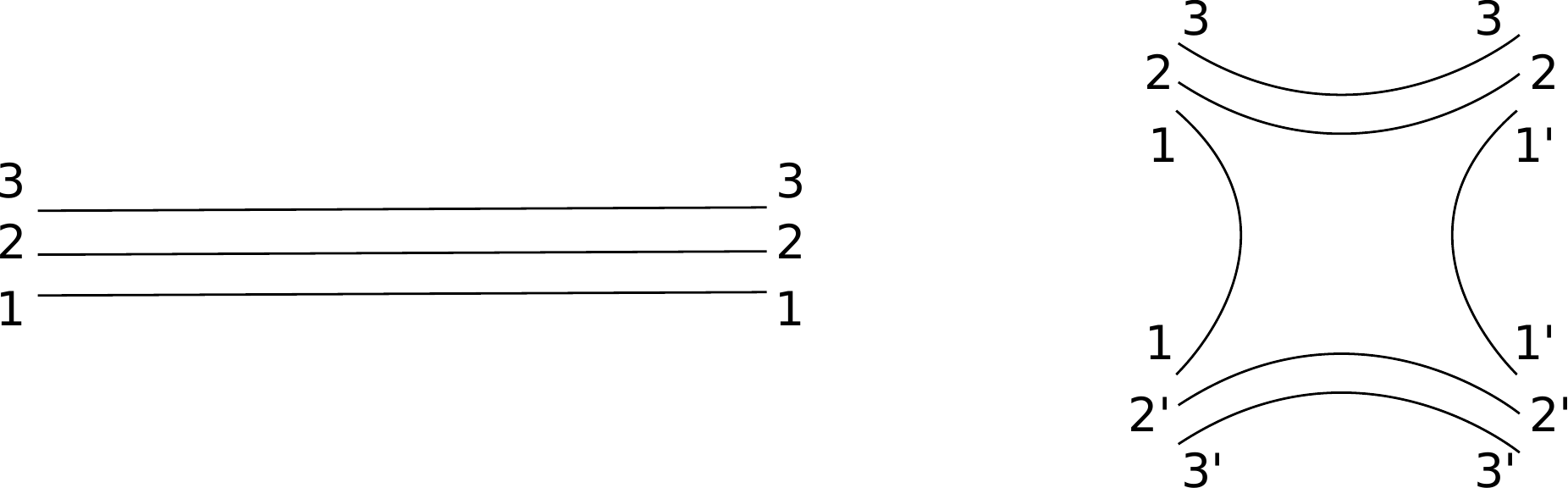} 
\caption{ {\small A rank $d=3$ propagator, as a stranded line (left), 
and the vertex $\Tr_{4;1}(\phi^4)$. }} 
\label{fig:feynman}
\end{minipage}
\end{figure}

We now pass at the quantum level. From \eqref{eq:3dkin}, we introduce a Gaussian field measure of covariance $C$ of the form
\bea
d\nu_C(\phi,\bar\phi) = \prod_{\bf P} d\phi_{\bf P}d\bar\phi_{\bf P}\,
e^{ -\Tr_2 [  \bar\phi \cdot ( K + \mu     )\cdot \phi] }\,,
\quad 
C = 1/(K+\mu)\,. 
\label{mesur}
\eea
Also called propagator, $C({\bf P};\bar{\bf P})$ is represented
graphically by a stranded line with $d$ segments (see an example
in rank 3 in Figure \ref{fig:feynman}). 
Let us treat the interaction part of the theory.  Tensor
field interactions are represented 
by stranded vertex graphs. For instance, associated with 
\eqref{choice1}, we obtain the vertex on the r.h.s of Figure \ref{fig:feynman}. As one observes, a $\phi^{2n}$ interaction may be
very well not symmetric with respect to its indices. At the level of the action $S^{\inter}$, we always sum over all colored symmetric terms
to be able to renormalize a theory. 

\begin{figure}[h]
\centering
     \begin{minipage}{.8\textwidth}
\includegraphics[angle=0, width=4cm, height=1.5cm]{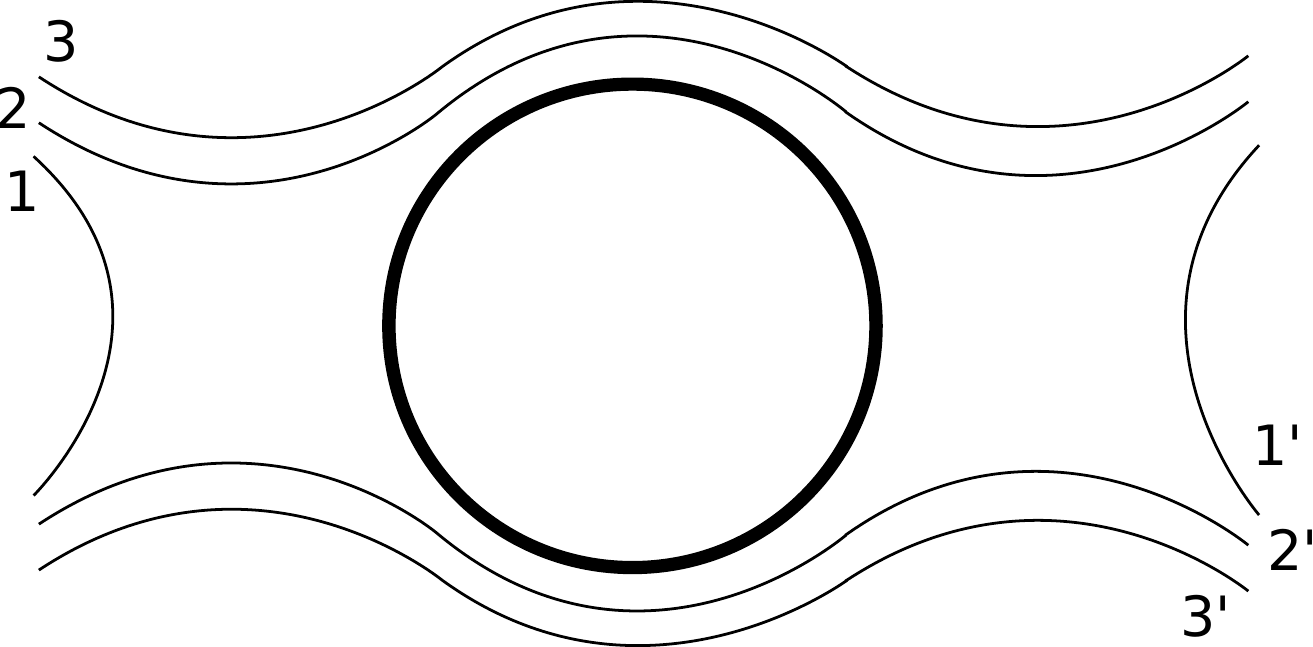} 
\caption{ {\small  An example of a rank $d=3$ TFT graph
and an internal face (put in bold) as a loop. }} 
\label{gra}
\end{minipage}
\centering
\end{figure}

As in any perturbative QFT, to study TFT correlators, we expand
them at small coupling constants, use the Gaussian measure $d\nu_C$ \eqref{mesur} to generate Feynman graph amplitudes via the ordinary Wick 
theorem. A Feynman graph, in the present instance, has a specific stranded structure and represent a simplicial complex in dimension $d$.
See an example in Figure \ref{gra}. In the most generic case, summing over infinite degrees of freedom implies divergent amplitudes. In TFTs, 
divergences are localized in the graphs by the presence of loops called internal faces (see again, Figure \ref{gra}).  

We consider now TFT with a field $\phi: (U(1)^D)^{\times d} \to \C$, 
producing a $D\times d$ field theory and seek conditions
to obtain a regularized and renormalizable TFT. 
Note that the two parameters $D$ and $d$  will play 
a different role. The kinetic term with  kernel $K$  can 
be extended to $(U(1)^D)^{\times d}$.

\

\noindent{\bf Renormalizable TFTs.}
The renormalization is performed via
a multiscale analysis.\cite{Rivasseau:1991ub} 
Such a program begins with a slice decomposition of the propagator as $C=\sum_{i=0}^{\infty} C_{i}$
where each propagator in the slice $i$, namely $C_i$ satisfies the 
upper bound
$C_i \leq    k M^{-2i} 
 e^{-\delta M^{-i}(\sum_{s=1}^d  |p_{s}|^{a} + \mu)} $
for some constants $k$, $M >1$ and $i>0$, and $C_0 \leq k$. 
Note that high $i$ should select large momenta $p_{s}$ of order $M^{\frac{i}{a}}$. We call this ultraviolet (UV) regime corresponding to short distances on $U(1)$. In opposite the regime, the slice $i=0$ refers to the infrared (IR). 
The regularization scheme requires to 
introduce UV cut-off $\Lambda$ on the sum over slices $i$
and so that the  regularized propagator is given by $C^\Lambda = \sum_{i=0}^{\Lambda} C_{i}$. 

An amplitude associated with a graph $\cG(\cV,\cL)$ expresses,
as in the usual way, as a product of  propagator lines and vertex  operators: 
$A_{\cG}=\sum_{p_{s;v} } \,\prod_{\ell \in \cL}
C[\{{\bf P}_{v(\ell)}\},\{{\bf P}'_{v'(\ell)}\}] \prod_{v\in \cV;s}
\delta_{p_{s;v};p'_{s;v}}$. We perform a slice decomposition of all propagators, and collect the momentum scales $i_\ell \in [0, \Lambda]$ in 
a multi-index $m=(i_{\ell})_{\ell \in\cL}$ called momentum attribution.
Then, we write $A_{\cG}= \sum_{m} A_{\cG;m}$, where $A_{\cG;m}$
is the amplitude at fixed momentum attribution. 
The question is to provide the behavior of $A_{\cG;m}$ after an optimal integration of internal momenta in terms of the parameters $M$ and 
of the set of the so-called quasi local subgraphs $\{G^i_k\}_{i,k}$.\cite{Rivasseau:1991ub}

The following statement holds
(power counting theorem):\cite{rankd} 
Let $\cG$ be a connected graph with set $\cL(\cG)$ of lines
of size $L(\cG)$, 
and set $\cF_{\inter}(\cG)$ of internal faces of size $F_{\inter}(\cG)$, then there exists
a constant $K_{\cG}$ depending on the graph such that 
\beq\label{degdiv}
|A_{\cG, m}| \leq  K_{\cG} \prod_{(i,k)\in \N^2} M^{\omega_{\rm d}(G^i_k)}\,,
\qquad 
\omega_{\rm d}(G^i_k) = -2aL(G^i_k)  +D\,F_{\inter}(G^i_k) \,.
\eeq 
The superficial divergence degree  $\omega_{\rm d}(\cG)$ of the graph $\cG$ determines if the amplitude associated with $\cG$ is divergent (when $\omega_{\rm d}(\cG)\geq 0$) or not.

As a second stage, we must treat the divergence degree and
express the number of internal faces in terms
of  Gurau's degree of the underlying colored graph \cite{largeN}  and
of the degree of the boundary graph. The boundary graph  encodes the boundary of the dual simplicial complex. 
As a definition, Gurau's degree is a sum of genera of 
canonical-colored surfaces of the TFT graph. 
It is proved that the amplitude is maximally divergent if
it underlying graph has a vanishing degree. 
Studying the degree of divergence, it appears that the set
of diverging graphs includes those with a vanishing degree, with a vanishing degree of their boundary graph and a restricted number of external fields. One obtains conditions on $(a,D,d)$ and the maximal 
valence $k_{\max}$ in the vertex interactions yielding a renormalizable model. A subtraction scheme of the divergences can then be identified.
The equations of the renormalized couplings in terms 
of the initial couplings define the so-called $\beta$-function equations which encode the renormalization group flow of the model. 
We obtain the table 1 of renormalizable models 
as well as their UV asymptotic behavior after calculation of their $\beta$-functions.\cite{rankd} 
{\footnotesize{\begin{table}[h]
\begin{center}
\begin{tabular}{lccccc|cc|}
\hline\hline
Type& $G_D$ & $\Phi^{k_{\max}}$  &  $d$ & $a$ & Renormalizability &   UV behavior \\
\hline\hline
TFT & $U(1)$ &   $\Phi^{6}$ & 4 & 1 & Just- &  ? \\
TFT & $U(1)$ &   $\Phi^{4}$  & 3 & $\frac12$ & Just- & AF\\
TFT & $U(1)$ & $\Phi^{6}$ & 3 &  $\frac23$ & Just- & ? \\
TFT & $U(1)$ & $\Phi^{4}$ & 4 & $\frac34$ & Just- &  AF \\
TFT & $U(1)$ & $\Phi^{4}$ & 5 &  1 & Just- &  AF \\
TFT & $U(1)^2$ & $\Phi^{4}$ & 3 & 1 & Just- &  AF \\
TFT & $U(1)$ & $\Phi^{2k}$ & 3 & 1 & Super- & AF\\
\hline
\end{tabular}\end{center}
\end{table}
}}

\section{Conclusion}
\label{ccl}

We have identified a set of renormalizable actions built with tensor fields.
At the UV-limit, many models turned out to be asymptotically free,
in particular $\phi^4$ models. The UV-behavior of the $\phi^6$ models
 is more subtle. 
The simplest $\phi^6$ TFT model
has been initially claimed AF\cite{josephBeta} but there are  indications that this model could be actually safe in the UV.\cite{Sylvain}  The existence of renormalizable tensor actions actually goes beyond the scope of the models presented in section \ref{sect:TFTs} where TFTs appear in their simplest form, 
see for instance.\cite{Carrozza:2013mna} For the lack of space,
we cannot review these models in details.

Having understood their small coupling behavior, the 
study of TFTs has been recently pursued at the nonperturbative 
level through the Functional Renormalization Group approach.\cite{Benedetti:2014qsa} As a new result, the existence of an IR fixed point 
seems to be generic in TFTs. If confirmed, an IR fixed point
also hints  at a phase transition.  In analogy with usual complex scalar field theory, the likely phases of TFT will be described by a spontaneous symmetry breaking mechanism. The two phases, the symmetric and
broken one will correspond to 
positive and negative mass, respectively. The broken phase might
be associated with a new condensed and geometrical ground state.\cite{GFTcondensate} 
This point deserves full investigations.

\section*{Acknowledgments}
The author acknowledges the support of the Max-Planck Institute 
for Gravitational Physics, Albert Einstein Institute and
warmly thank all the organizers of ICMP2015, in particular, Prof. R. Benguria, for their invitation and memorable welcome in Chile.

\end{document}